\documentclass[twocolumn]{aastex631}
\usepackage{graphicx,lipsum, booktabs, footnote}
\usepackage{wrapfig}
\usepackage{float}
\usepackage{multirow}
\usepackage{graphicx, subfigure}

\usepackage[autostyle]{csquotes}

\begin{document}

\title{Globular Clusters GMRT Pulsar Search (GCGPS) II: Discovery of five MSPs in M69 and M70}

\author[0009-0006-7995-5871]{Jyotirmoy Das}
\author[0000-0002-2892-8025]{Jayanta Roy}
\affiliation{National Centre for Radio Astrophysics (NCRA) \\
Pune-411007, Maharashtra, India}

\author[0000-0003-1307-9435]{Paulo C. C. Freire}
\affiliation{Max-Planck-Institut für Radioastronomie (MPIfR) \\
 Auf dem Hügel 69-53121, Bonn, Germany}
 
\author[0000-0001-5799-9714]{Scott Ransom}
\affiliation{National Radio Astronomy Observatory (NRAO) \\
 Charlottesville, Virginia, United States}

\author[0000-0002-6287-6900]{Bhaswati Bhattacharyya}
\affiliation{National Centre for Radio Astrophysics (NCRA) \\
Pune-411007, Maharashtra, India}

\author[0000-0003-2797-0595]{Karel Adamek}
\affiliation{Department of Physics, Silesian University in Opava \\
Opava, 74601, Czech Republic}

\author[0000-0003-1756-3064]{Wes Armour}
\affiliation{Oxford e-Research Centre (OeRC) \\
University of Oxford, Oxford-OX13PJ, United Kingdom}

\author[0000-0002-6631-1077]{Sanjay Kudale}
\affiliation{National Centre for Radio Astrophysics (NCRA) \\
Pune-411007, Maharashtra, India}
\affiliation{Giant Metrewave Radio Telescope (GMRT) \\
Khodad, Pune- 410504, Maharashtra, India}

\author[0009-0008-1233-6915]{Mekhala V. Muley}
\affiliation{Giant Metrewave Radio Telescope (GMRT) \\
Khodad, Pune- 410504, Maharashtra, India}

\begin{abstract}
This paper reports recent discoveries from the Globular Clusters GMRT Pulsar Search (GCGPS) survey, which aims to uncover pulsars in the globular clusters (GCs) of the Milky Way using the upgraded Giant Metrewave Radio Telescope (uGMRT). Utilising the Band-4 (550$–$750 MHz) and Band-3 (300$–$500 MHz) receivers, the survey targets GCs accessible to uGMRT ($-53^\circ\,<\,\delta\,<\,-17^\circ$), excluding the declination range that can be covered by the Five-hundred-meter Aperture Spherical radio Telescope (FAST). The survey focuses on GCs that have not previously been searched with comparable sensitivity in these radio frequencies. In this paper, we present the discovery of five MSPs in two GCs, $-$ NGC~6637 (Messier69/ M69) and NGC~6681 (Messier70/ M70), each hosting MSPs identified here for the first time. Observations of M69 led to the discovery of two MSPs: J1831$-$3220A (M69A) and J1831$-$3220B (M69B), both of which we localize with arcsecond precision using interferometric imaging. Observations of M70 resulted in three new MSPs: J1843$-$3217A (M70A), J1843$-$3217B (M70B), and J1843$-$3217C (M70C). Although direct imaging did not yield precise localizations for these MSPs, we provide initial estimates based on uGMRT beam forming and imaging analysis. Additionally, we present preliminary imaging results for other observed GCs, and in cases of non-detections, we report upper limits on pulsed emission based on the rms noise levels in the image plane.

\end{abstract}

\keywords{Radio astronomy; LMXBs; Pulsar survey: GCGPS; Globular Clusters: NGC~6637, NGC6681; Pulsars: J1831$-$3220A, J1831$-$3220B, J1843$-$3217A, J1843$-$3217B, J1843$-$3217C}

\section{Introduction}
\label{sec:intro}

The close interaction between an old neutron star (NS) and a low-mass star can lead to the formation of a low-mass X-ray binary (LMXB) system \citep{Kalogera_1996, Kalogera_1998_1, Kalogera_1998_2}. In such systems, the NS accretes matter from its companion, gradually spinning up in the process. The high stellar densities and interaction rates found in globular clusters (GCs) \citep{Verbunt_Hut_1987, Pooley_2003, Bahramian_2013} make them ideal environments for the dynamical formation of LMXBs—an occurrence that is rare in the Galactic disk. As a result, GCs host approximately $10^3$ more LMXBs per unit mass than the Galactic disk \citep{Sarazin_2003}. Over time, these LMXBs evolve into pulsars with spin periods of just a few milliseconds, known as recycled or millisecond pulsars (MSPs); generally with spin periods $P_\mathrm{s} < 20$ ms \citep{Becker_1999} and low magnetic fields ($B \sim 10^8 - 10^{10} \, \rm G$), usually accompanied by a low-mass white dwarf ($M_\mathrm{\rm WD} < 0.20\, \rm \:M_\mathrm{\odot}$, see \citealt{Istrate_2014}) or a non-degenerate companion \citep[for a review, see][]{Tauris_2023}. The abundance of LMXBs in GCs leads to a correspondingly high population of MSPs: of the $>$700 known MSPs, $\sim$ 340 are located in GCs\footnote{Based on data from the ATNF pulsar catalogue, \url{https://www.atnf.csiro.au/research/pulsar/psrcat/} \citep{Manchester_2005} and Paulo Freire's GC pulsar list, see \url{https://www3.mpifr-bonn.mpg.de/staff/pfreire/GCpsr.html}; the total number of known pulsars in GCs is currently 359.}.

The high stellar densities of GCs mean that some MSP binaries have properties unlike any found in the Galactic disk, such as high orbital eccentricities and massive companions. These binaries result from exchange encounters that happen after the pulsar is recycled. 
These unusual characteristics make them, potentially, powerful probes for studying gravity in the strong-field regime. For instance, J0514$-$4002E in NGC~1851 might be an MSP–black hole system \citep{Barr_2024}, which could allow tests of alternative gravity theories beyond the current capabilities of binary pulsar timing \citep{Freire_Wex_2024}.

These rare systems highlight the importance of continued pulsar searches in GCs. This is one of the main motivations of the Globular Clusters GMRT Pulsar Search (GCGPS) survey. This survey, including a more complete motivation, target selection and technical details, is discussed in detail by \cite{Das_2025}, which is Paper I in this series; it is also the designation we will use throughout this work.

In this work (Paper II), we provide an update on the current status of the survey and its data analysis in Section~\ref{sec:surveystatus}. In Section~\ref{sec:discoveries}, we report the discovery of five MSPs in NGC~6637 (M69) and NGC~6681 (M70). In Section~\ref{sec:GCimaging}, we present the radio imaging analysis of the GCGPS-observed clusters, which helps localize the newly discovered pulsars, and provides a set of unassociated faint radio sources, worth following up for potential new MSPs, along with flux-density upper limits for the non-detections. The final imaging data (in FITS format) associated with this paper are also made publicly available (DOI: \href{https://doi.org/10.5281/zenodo.17781807}{10.5281/zenodo.17781807}), offering a valuable resource for planning more targeted and sensitive follow-up searches. Finally, we summarise our findings in Section~\ref{sec:summary}.

\section{OBSERVATIONS AND SURVEY STATUS}
\label{sec:surveystatus}

The GCGPS observations began in May 2023. In each half-yearly uGMRT cycle, we observe a subset of the selected globular clusters to search for new MSPs. Each GC is observed using two simultaneously formed Phased Array (PA) beams, one formed with the central square (CSQ) antennas and the other using antennas up to the third arm, in either Band-3 or Band-4, depending on the target. This configuration ensures the best sensitivity possible with the uGMRT across both the full cluster and its core. For GCs with known MSPs, we record coherently dedispersed (CD) phased array data at median GC DM with a time resolution of 40.96 $\mu$s and 512 frequency channels across a 200 MHz bandwidth. For GCs with no known MSPs a priori, we record standard phased array(PA) data with 81.92 $\mu$s time resolution and 4096 channels, without any coherent dedispersion. Alongside beam data, we also record the interformatric data to image the GC field for further analysis. The data analysis procedures followed here are the same as those described in detail in Paper I.

Of the 31 GCs initially targeted in Phase I (listed in Table 2 of Paper I), we have to date completed observations and data analysis for one GC in Band 3 and 21 GCs in Band 4 (see Table~\ref{table:observation}). Although NGC 5897 was listed as a Band 3 target in Table 2 of Paper I, it was observed in Band 4 instead due to feed-rotation constraints, and its Band 4 analysis is therefore presented in this paper. Based on this progress, approximately 50\% of the planned Phase I survey is complete in Band-3, and about 73\% is complete in Band 4. So far, this survey has led to the discovery of seven millisecond pulsars (MSPs) in four globular clusters where no pulsars were previously known\footnote{See the survey webpage: \url{http://www.ncra.tifr.res.in/~jroy/GC.html}}. The first discovery of this survey, J1617$-$2258A in NGC~6093, along with the survey details and its timing follow-up, was presented in Paper I. In this paper, we report five additional discoveries in NGC~6637 (Messier69/ M69) and NGC~6681 (Messier70/ M70), present their initial study and imaging analysis of all observed and analysed GCs.

Apart from the globular clusters (GCs) observed and analysed so far in this survey (a subset of the Phase-I targets), Table~\ref{table:observation} lists their observational details. The remaining Phase-I observations, as well as the planned Phase-II observations—which will include several GCs previously excluded in Phase I due to sensitivity limits or restricted sky coverage—are currently ongoing. These Phase-II observations are being carried out using the newly commissioned multibeam SPOTLIGHT\footnote{The SPOTLIGHT website: \url{https://spotlight.ncra.tifr.res.in/}}\citep{Roy_2024} backend, which provides significantly higher sensitivity along with multibeam coverage (currently $\sim$160 beams), substantially exceeding the single-pointing capability of the Phase-I GCGPS survey. The MeerKAT’s TRAPUM \citep{Ridolfi2022, Abbate2022} survey, currently the most prominent survey in the Southern hemisphere\footnote{The survey website: \url{https://www.trapum.org/}}, has greatly benefited from the large number of beams provided by a similar beamformer installed at the MeerKAT telescope, the Filterbanking Beamformer User Supplied Equipment \citep[FBFUSE,][]{Barr_2018,Chen_2021}. This has been especially useful for surveys of extended globular clusters \citep{Chen_2023}.

A full-array 30 antenna phased uGMRT Band-4 PA beam is expected to provide a theoretical sensitivity roughly $\sim$2 times higher than the scaled L-band sensitivity of TRAPUM (see Section 3.3 of Paper I for details). This improvement in both sensitivity and sky coverage will allow us to observe targets that were earlier excluded because of limited coverage or comparatively lower sensitivity. The results from these ongoing and upcoming observations will be presented in future publications from this survey.

\begin{table*}
\vspace{-10pt}
\centering
\caption{The list of GCs observed and analysed so far under the GCGPS survey. The \enquote{Obs epoch} is the Modified Julian Date (MJD) at which the associated GC was observed. The uGMRT bands used under the GCGPS survey are Band-3 (330$-$500 MHz) and Band-4 (550$-$650 MHz), and the associated band for each GC observation is mentioned in the column titled \enquote{uGMRT BAND}. The PA and CD in the \enquote{Obs mode} stand for the \enquote{Phased Array} and \enquote{Coherently Dedispersed} mode, whichever was used to take the data, as discussed in Section \ref{sec:surveystatus}. For all the observations, the beam was formed at the GC centre and the GC centre RA (J2000) and Dec (J2000) values were taken from the W. E. Harris list of GC parameters, \citep[2010 edition]{Harris_1996}. $\rm S^{c}_\mathrm{min}$ is the continuum 10$\sigma$ and $\rm S^{p}_\mathrm{min}$ is the PA/CD beam data 10$\sigma$ detection limit for each obsevation (see Section \ref{subsec:nondetection}).}
\begin{tabular}{|c|c|c|c|c|c|c|c|c|}
\toprule
\midrule
GC name & Obs epoch & uGMRT BAND & Obs mode & On-source time & Beam center (RA, Dec) & $\rm S^{c}_\mathrm{min}$ & $\rm S^{p}_\mathrm{min}$\\
(NGC XXXX) & MJD & (BAND-4/3) & (PA/CD) & (sec) & (hh:mm:ss, dd:mm:ss) & ($\rm \mu Jy$) & ($\rm \mu Jy$)\\
\midrule
NGC~1904 & 60282 & BAND-4 & PA & 9060 & (05 24 11.09, $-$24 31 29.0) & 112 & 71\\
\midrule
NGC~4590 & 60273 & BAND-4 & PA & 8220 & (12 39 27.98, $-$26 44 38.6) &  413 & 270\\
\midrule
NGC~5286 & 60608 & BAND-4 & PA & 6960 & (13 46 26.81, $-$51 22 27.3) & 361 & 236 \\
\midrule
NGC~5897 & 60447 & BAND-4 & PA & 6960 & (15 17 24.50, $-$21 00 37.0) & 201 & 131\\
\midrule
NGC~5986 & 60110 & BAND-4 & CD & 4560 & (15 46 03.00, $-$37 47 11.1) & 173 & 117\\
\midrule
NGC~6093 & 60126 & BAND-4 & PA & 8220 & (16 17 02.41, $-$22 58 33.9) & 132 & 86\\
\midrule
NGC~6139 & 60372 & BAND-4 & PA & 7560 & (16 27 40.37, $-$38 50 55.5) & 1553 & 869\\
\midrule
NGC~6293 & 60664 & BAND-4 & PA & 7200 & (17 10 10.20, $-$26 34 55.5) & 802 & 544\\
\midrule
NGC~6333 & 60447 & BAND-4 & PA & 7440 & (17 19 11.26, $-$18 30 57.4) & 186 & 117\\
\midrule
NGC~6355 & 60632 & BAND-4 & PA & 8580 & (17 23 58.59, $-$26 21 12.3) & 532 & 348\\
\midrule
NGC~6388 & 60066 & BAND-4 & PA & 5400 & (17 36 17.23, $-$44 44 07.8) & 146 & 92\\
\midrule
NGC~6528 & 60664 & BAND-4 & PA & 7260 & (18 04 49.64, $-$30 03 22.6) & 353 & 231\\
\midrule
NGC~6541 & 60357 & BAND-4 & PA & 9660 & (18 08 02.36, $-$43 42 53.6) & 261 & 171\\
\midrule
NGC~6553 & 60632 & BAND-4 & PA & 7560 & (18 09 17.60, $-$25 54 31.3) & 98 & 62\\
\midrule
NGC~6569 & 60981 & BAND-4 & PA & 7860 & (18 13 38.80, $-$31 49 36.8) & 120 & 78\\
\midrule
NGC~6637 & 60447 & BAND-4 & PA & 6960 & (18 31 23.10, $-$32 20 53.1) & 92 & 62\\
\midrule
NGC~6638 & 60981 & BAND-4 & PA & 9660 & (18 30 56.10, $-$25 29 50.9) & 72 & 45\\
\midrule
NGC~6642 & 60379 & BAND-4 & PA & 9000 & (18 31 54.10, $-$23 28 30.7) & 125 & 85\\
\midrule
NGC~6652 & 60632 & BAND-4 & CD & 7080 & (18 35 45.63, $-$32 59 26.6) & 241 & 163\\
\midrule
NGC~6681 & 60431 & BAND-4 & PA & 8640 & (18 43 12.76, $-$32 17 31.6) & 212 & 112\\
\midrule
NGC~6723 & 60146 & BAND-4 & PA & 8160 & (18 59 33.15, $-$36 37 56.1) & 456 & 277\\
\midrule
NGC~6809 & 60421 & BAND-3 & PA & 13560 & (19 39 59.71, $-$30 57 53.1) & 511 & 322\\
\bottomrule
\end{tabular}
\label{table:observation}
\end{table*}

\section{DISCOVERIES}
\label{sec:discoveries}

We now report the discovery of five MSPs in the PA beam data of two GCs with no previously known pulsars and provide a detailed summary of their initial parameters. 

\subsection{Discoveries in NGC~6637 (M69)}
\label{subsubsection:M69discovery}
As shown in Table \ref{table:observation}, we observed M69 in May 2024 for two hours of on-source time as part of the GCGPS survey. As described in Section \ref{sec:surveystatus}, two simultaneous uGMRT Band-4 phased-array (PA) beams: the CSQ beam and the 3rd-arm beam, were formed to observe M69 with optimal sensitivity for pulsar discovery. We adopted the predicted dispersion measure of the YMW16 model ($\rm DM^* = 119.8\,pc\,cm^{-3}$, see \cite{Yao_2017}) and searched a DM range of 40 to 180 $\rm pc\,cm^{-3}$ (following the DM search strategy detailed in Paper I) with a DM step of 0.1 $\rm pc\,cm^{-3}$. An initial {\tt PRESTO}\footnote{GitHub link of ${\tt PRESTO}$ repository: \url{https://github.com/scottransom/presto}}\citep{Scott_2011} based acceleration search with a maximum 200 bin-drift correction ($Z_\mathrm{max} = \pm\,200$) was performed on the CSQ data using our CPU-based pipeline PSS (available at \url{https://github.com/jyotirmoydas5392/Pulsar_Search_Script}). This search led to the discovery of J1831$-$3220A, hereafter referred to as M69A. With a spin period of $P_\mathrm{s} \sim 3.81;\rm ms$, this pulsar was detected at a DM of 82.1 $\rm pc\,cm^{-3}$; a subsequent finer search refined this value to $\rm DM_{M69A} = 82.12\:\rm pc\,cm^{-3}$. The discovery signal-to-noise ratio (SNR) from the {\tt PRESTO PREPFOLD} folding of the CSQ data was 17.5$\sigma$.

Following this detection, we performed a more sensitive search to identify lower-SNR candidates, covering the DM range $\rm DM_{M69A} \pm 15\:pc\,cm^{-3}$ with a finer DM step of 0.02 $\rm pc\,cm^{-3}$. This search resulted in the detection of a second candidate, which was later confirmed in a follow-up observation as J1831$-$3220B, hereafter M69B. The discovery SNR of this pulsar in the CSQ data was approximately 10$\sigma$, with a spin period of $P_\mathrm{s} \sim 4.80\:\rm ms$ and a measured DM of 81.92 $\rm pc\,cm^{-3}$.

Having 3rd arm beam central theoretical sensitivity $\rm \sim\:1.57\:(N_\mathrm{2}/N_\mathrm{1}\:=\:22/14)$ times than the CSQ beam central sensitivity, a general acceleration search over the DM space $\rm DM_\mathrm{M69A}\:\pm\:15$ $\rm pc\,cm^{-3}$ with a DM step 0.02 $\rm pc.cm^{-3}$ was also performed on the third arm data for pulsars situating close to the GC center, but no promising candidates were detected.

To obtain the initial positional estimates, we folded the 3rd arm data for these two MSPs. The detection significance in both cases was reduced by a factor of 40\%$-$60\% compared to the CSQ detection, suggesting that the MSPs are located away from the cluster center, where the CSQ beam has 40\%$-$60\% higher sensitivity than the 3rd arm beam. Based on this estimate, we performed imaging of the field to localize the MSPs; this is discussed in Section~\ref{subsec:localization}.

During the discovery observation, M69A was detected with a significant acceleration ($P_{\mathrm{dot}}\sim 10^{-12}$), confirming that it's a binary MSP in a relatively compact orbit. Now, for M69B, we could not detect any significant acceleration, indicating that it is an isolated or an MSP in a binary orbit wide enough so that two hours of on-source time is not enough to detect the orbital acceleration. Timing follow-up observations are underway to obtain precise timing solutions for these systems, which we plan to publish in a subsequent paper as part of this survey.

\subsection{Discoveries in NGC~6681 (M70)}
\label{subsubsection:M70discovery}
Similar to the M69 observation, M70 was observed in May 2024 as part of the GCGPS survey (observation details in Table \ref{table:observation}). Two simultaneous uGMRT Band-4 phased-array (PA) beams: the CSQ beam and the 3rd-arm beam, were formed to observe M70 with optimal sensitivity for pulsar discovery. Using the YMW16 model predicted dispersion measure ($\rm DM^* = 105.2\,pc\,cm^{-3}$), and following the same search strategy as for M69, we performed an acceleration search with $Z_\mathrm{max}\, =\,\pm\,200$, over a DM range of 30–160 $\rm pc\,cm^{-3}$ with a DM step of 0.1 $\rm pc\,cm^{-3}$. The search on the CSQ data led to the discovery of three millisecond pulsars (MSPs) in this globular cluster. PSR J1843$-$3217A, hereafter M70A, an MSP with a spin period of $P_\mathrm{s} \sim 3.93,\rm ms$, was detected at a DM of $\sim70.9\:\rm pc\,cm^{-3}$ with an SNR of $\sim23.67,\sigma$. The strongest detection was from J1843$-$3217B (M70B; $P_\mathrm{s} \sim 4.44,\rm ms$), identified at a DM of $71.2\:\rm pc\,cm^{-3}$ with an SNR of about $34.55,\sigma$. The faintest of the three, J1843$-$3217C (M70C; $P_\mathrm{s} \sim 6.06,\rm ms$), was detected at a DM of $\sim71.1\:\rm  pc\,cm^{-3}$ with an SNR of $12.06,\sigma$ in the CSQ data. Following these detections, more precise DM values for each MSP were determined using targeted {\tt PRESTO}-based searches around the initial detection DMs; the refined values are listed in Table \ref{table:MSPdetails}.

All these MSPs were re-detected in the simultaneously recorded 3rd arm data with increased SNR. For all of them, the increment in SNR is about 40\%$-$60\%, confirming the MSP discovery. This also suggests that they are situated very close to the GC center, and inside the HPBW of the 3rd arm PA beam. Detailed discussion on the localization is done in Section \ref{subsec:localization}. Similar to M69, we also performed a acceleration search on the 3rd arm data over $\rm DM^*\:\pm\:15\, pc\,cm^{-3}$ with DM step = 0.02\,$\rm pc\,cm^{-3}$, to detect the MSPs that are very close to the GC center, and might have been missed in the less sensitive CSQ search, but couldn't detect any promising candidates.

\begin{figure*}
\centering
    \includegraphics[width=\textwidth]{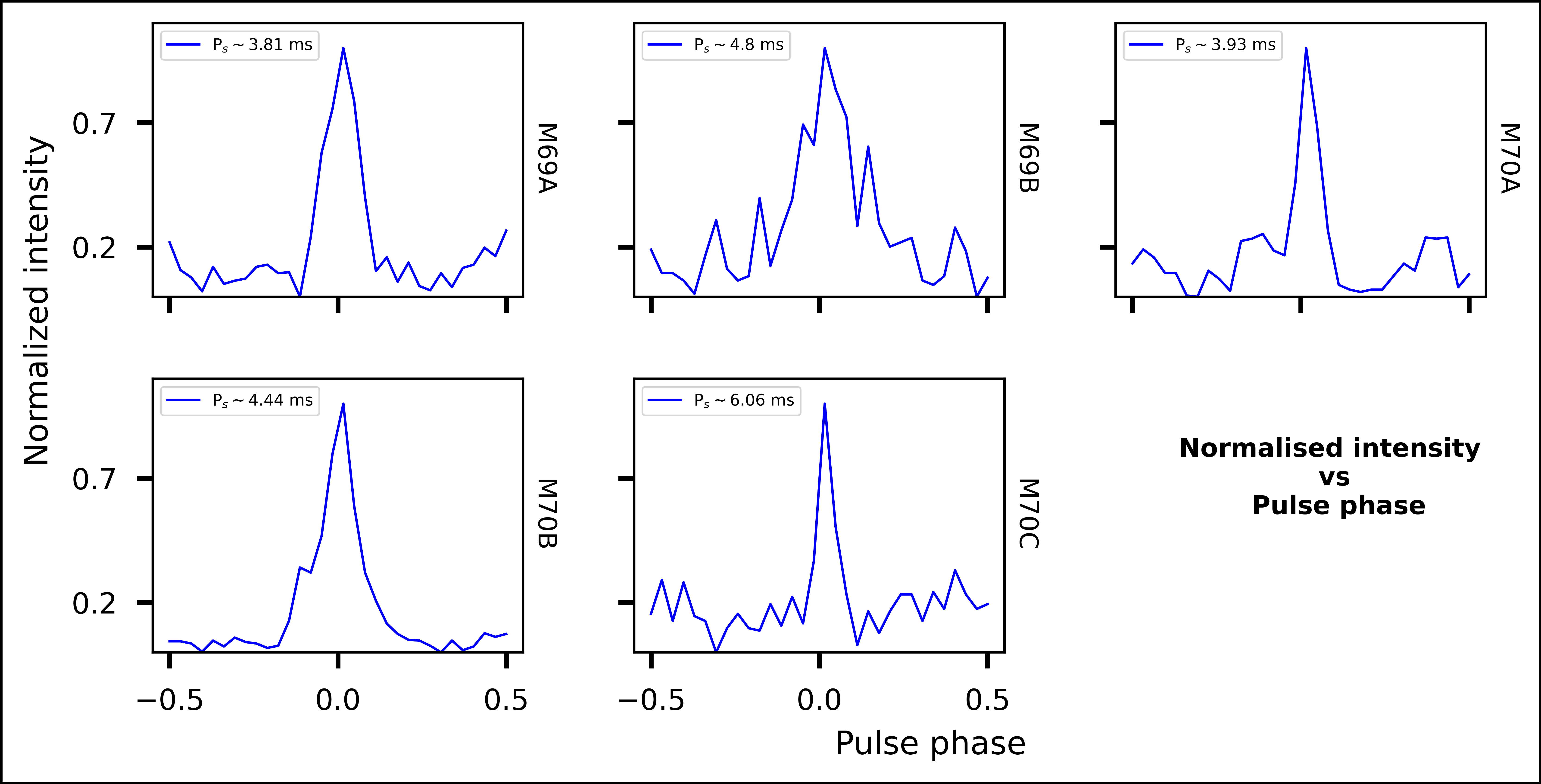}
    \caption{The normalised pulse profiles at uGMRT Band-4 (550$-$750 MHz) for the five MSPs discovered in two globular clusters as part of the GCGPS survey are shown. The first two profiles (left to right) are the M69A and M69B from NGC~6637 (Messier69/ M69), and the remaining (in the same order) are the M70A, M70B, and M70C from NGC~6681 (Messier70/ M70).}
    \label{fig:NewMSPsplot}
\end{figure*}

Including the two MSPs in M69, the normalized pulse profiles of all five MSPs are shown in Figure~\ref{fig:NewMSPsplot}, with their corresponding parameters listed in Table~\ref{table:MSPdetails}. In addition, the $\tt{PRESTO\:PREPFOLD}$ outputs for all five MSPs are provided in the Appendix (Section~\ref{appendix:MSP_prepfold_outputs}, Figure \ref{fig:GCGPS_MSP_folded_profiles}).

\begin{figure}
\centering
    \includegraphics[width=\columnwidth]{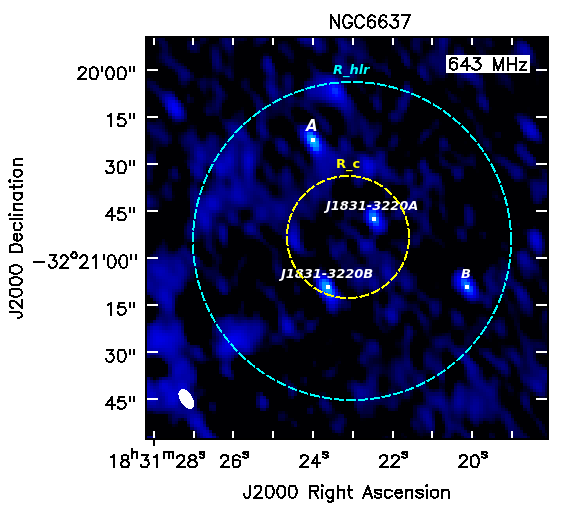}
    \vspace{-30pt}
    \caption{The uGMRT Band-4 radio image of the GC field M69, generated using the automated pipeline {\tt CAPTURE}. The image shows the localized positions of the two newly discovered millisecond pulsars (MSPs), J1831$-$3220A (M69A) and J1831$-$3220B (M69B), along with the marked core and half-light radius of the cluster. The sources \enquote{A} and \enquote{B} are two probable MSP candidates.}
    \label{fig:NGC6637_image}
\end{figure}

\begin{figure}
\centering
    \includegraphics[width=\columnwidth]{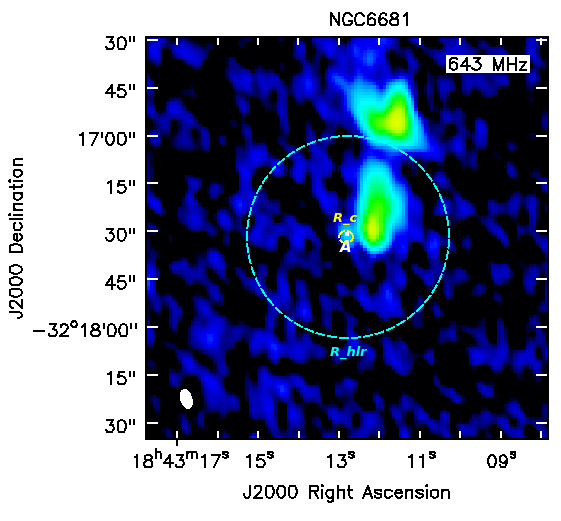}
    \caption{The {\tt CAPTURE} generated uGMRT Band-4 radio image of the GC field M70, where we can see the bright extended source close to the GC center, which prevented us from localizing the MSPs in the image plane.}
    \label{fig:NGC6681_image}
\end{figure}

\begin{table*}
\centering
\caption{This table presents the details of five millisecond pulsars (MSPs) discovered as part of the GCGPS survey. Two of these MSPs are located in NGC~6637 (Messier69/ M69), while the remaining three are in NGC~6681 (Messier70/ M70). For each pulsar, key discovery parameters—including spin period, dispersion measure (DM), {\tt PRESTO PREPFOLD} signal-to-noise ratio (SNR), and initial localized position (where confirmed)-are provided based on the discovery observations, initial imaging, and beam simulation analysis. For the localized MSPs in the image plane, we also provide the measured flux density ($\rm S_{\mathrm \nu}$, at 650 MHz) from the image analysis. The unconfirmed results are marked with \enquote{TBC} (To Be Confirmed), wherever applicable in the table.}
\begin{tabular}{|c|c|c|c|c|c|c|}
\toprule
GC name & Pulsar & Period & DM & Discovery SNR & RA (J2000) and Dec (J2000) & $\rm S_{\mathrm \nu}$ (650 MHz) \\
(NGC XXXX) & & (ms) & ($\rm pc\,cm^{-3}$) & (CSQ, 3rd arm) & (hh:mm:ss, dd:mm:ss) & ($\mu Jy$) \\
\midrule
\multirow{2}{*}{NGC~6637} & M69A & 3.81 & 82.12 & ($17.5\sigma,\,11.2\sigma$) & (18:31:22.43, $-$32.20.47.36) &  224 $\pm$ 13\\
\cline{2-7}
     (Messier69/ M69) & M69B & 4.80 & 81.92 & ($10.0\sigma,\,7.6\sigma$) & (18:31:23.65, $-$32:21:10.31) & 278 $\pm$ 11 \\
\midrule
\multirow{3}{*}{NGC~6681} & M70A & 3.93 & 70.65 & ($34.3\sigma,\,47.8\sigma$) & TBC & TBC \\
\cline{2-7}
    & M70B & 4.44 & 71.14 & ($16.9\sigma,\,23.7\sigma$) & TBC & TBC \\
\cline{2-7}
    (Messier70/ M70) &  M70C & 6.06 & 71.06 & ($11.6\sigma,\,16.8\sigma$) & TBC & TBC \\
\bottomrule
\end{tabular}
\label{table:MSPdetails}
\end{table*}

\section{Imaging of the GCGPS observed GCs}
\label{sec:GCimaging}
From the simultaneously recorded interferometric data, we imaged all GCGPS-observed globular clusters using the automated pipeline {\tt CAPTURE}. The {\tt CAPTURE} pipeline is designed to process uGMRT interferometric data in an automated manner, provided the observations follow a standard imaging strategy. Originally, the pipeline required the presence of a dedicated standard flux calibrator to perform absolute flux calibration. However, for several GCGPS observations, we prioritised longer on-source integration time and therefore did not observe a standard flux calibrator. Instead, only bright phase calibrators were observed as part of the phasing operations. To enable imaging of these non-standard datasets, we modified the pipeline to perform flux calibration using the observed phase calibrators by manually supplying appropriate calibrator models. These models were constructed using the most recent multi-band observations (from GCGPS survey as well) or measurements (mostly from VLA calibrator database\footnote{The VLA calibrator database: \url{https://science.nrao.edu/facilities/vla/observing/callist}} ) available for the corresponding phase calibrators. The adopted flux densities and spectral indices were further validated using existing GCGPS observations that included standard 3C flux calibrators, with the same phase calibrator used as a common reference. This confirmed that the flux densities derived for the phase calibrators are consistent with their catalogued values. This approach eliminated the need for dedicated flux calibrator observations and enabled reliable imaging of the observed GC fields while maximising on-source integration time.

\subsection{localization of pulsars}
\label{subsec:localization}
As discussed in Section \ref{subsubsection:M69discovery} and \ref{subsubsection:M70discovery}, both GCs were observed simultaneously using the CSQ and 3rd arm PA beams during the discovery observations. Aided by beam multibeam (CSQ and the 3rd arm) detection, the imaging from the interferometric data allowed us to localize the MSPs within the image field efficiently, provided they were detectable as point sources.

Imaging of M69 revealed multiple radio sources within the field, some of which could correspond to the newly discovered MSPs. To identify the most likely candidates, we used the detection signal-to-noise ratio (SNR) between the CSQ and 3rd arm beams from the discovery observation, along with beam simulation and radiometer equation \citep{Dewey_1985} estimated fluxes. The third arm to CSQ SNR ratios for PSRs M69A and M69B were approximately 0.64 and 0.76, respectively. We simulated the beam responses for both the CSQ and third arm beams during the discovery observation using the uGMRT beam simulation tool {\tt uGMRT\_beam\_fov.py}. Based on the observed SNR ratios and the beam response model, we estimated the radial offsets of the MSPs to be within $\sim$15$–$30 arcseconds from the cluster center. Guided by these estimates, along with estimated fluxes, imaging revealed two promising point source candidates consistent with the expected positions of the MSPs, along with two additional point sources located slightly farther from the predicted radial range. In subsequent timing follow-up observations, each of the four candidate positions was targeted with two simultaneous beams centerd on their respective locations. The resulting third-arm-to-CSQ SNR ratios confirmed that the two inner candidates correspond to the newly discovered MSPs in M69. The localized positions of these two MSPs and their associated radio continuum sources are shown in Figure~\ref{fig:NGC6637_image}, and their RA, Dec, and measured flux densities from the images are listed in Table~\ref{table:MSPdetails}. The two outer candidates (marked as \enquote{A} and \enquote{B} in Figure~\ref{fig:NGC6637_image}), which were not associated with the discovered MSPs or any other known pulsars, remain strong candidates for potential MSPs.

For the three MSPs in M70 (M70A, M70B, and M70C), the third arm to CSQ detection SNR ratios were approximately 1.39, 1.49, and 1.45, respectively. Similar to our analysis for M69, we simulated the uGMRT beam responses (CSQ and third arm) at the time of the discovery observation for M70 and used the observed SNR ratios to estimate the likely radial positions of the MSPs within the field. The simulations suggest that the expected positions of all three MSPs lie within $\sim$10 arcseconds of the cluster center, with the highest probability density concentrated at the very core. Imaging of M70 revealed a relatively strong, extended radio source located near the cluster center, which is unlikely to be associated with any of the MSPs due to its morphology and high flux density. In addition to this, a fainter, compact point source was detected almost exactly at the cluster center, marked as \enquote{A} in Figure~\ref{fig:NGC6681_image}. Given its compactness and alignment with the predicted positions, this source is a promising candidate for any of the three MSPs. No other plausible sources were identified within the cluster region. However, dedicated timing follow-up is required to confirm their actual positions and to establish a definitive association with the imaged source.

\subsection{The GC images}
\label{subsec:GC_images}
We present all twenty-two beam-processed GC images in the Appendix section, grouped into three sets (see Section \ref{appendix:GC_images}, Figures~\ref{fig:GCGPS_clusters_1}, \ref{fig:GCGPS_clusters_2}, and \ref{fig:GCGPS_clusters_3}). These images reveal a wealth of information about each GC field, providing a valuable basis for extracting further scientific insights.

The clusters NGC~5986 and NGC~6652 had known MSPs prior to the GCGPS observations. In the case of NGC~5986, the imaging revealed several faint compact sources (with flux densities of a few hundred $\mu$Jy) located outside the half-light radius, which could be potential MSP candidates. However, the lack of a well-localised position for the previously discovered pulsar in this cluster (J1546$-$3747A) did not allow us to associate it with any of the detected sources in the image field. For NGC~6652, similar to the findings of \cite{Gautam2022}, we detect the source that they associated with J1835$-$3259B (marked in the image), but no findings for J1835$-$3259A, along with several other faint compact sources that are not associated with any known radio emitters. These additional detections are promising candidates for follow-up studies aimed at identifying new MSPs.

There are now four GCs (NGC~1904, NGC~6093, NGC~6637, and NGC~6681) that host first known MSPs for each GC, discovered through the GCGPS survey. For NGC~1904, the localisation work for the GCGPS-discovered pulsar NGC~1904A (J0524$-$2431A) is currently ongoing. We include the image of NGC~1904 in this paper for completeness. In NGC~6093, three MSPs are currently known: the first discovered by the GCGPS survey (PAPER~I), and two subsequently found by the TRAPUM survey\footnote{See \url{https://www3.mpifr-bonn.mpg.de/staff/pfreire/GCpsr.html}}. Of these, only NGC~6093A (J1617$-$2258A) has a well-localised position (from PAPER~I), which we mark in the image. Similarly, for NGC~6637, the localised MSPs discovered in this work are marked in its image. The image of NGC~6681 is also presented, although the positions of the MSPs discovered by TRAPUM in this cluster (three more after the first three discoveries from the GCGPS survey) are not yet publicly available. Across all four clusters, the images reveal several faint yet compact, unassociated radio sources. These may correspond to the GCGPS-discovered MSPs (for NGC~1904 and NGC~6681), or TRAPUM-discovered MSPs (in NGC~6093 and NGC~6681), whose positions are not yet released, or they could represent entirely new, yet-undiscovered MSPs. These sources warrant targeted follow-up observations.

Except for these six GCs, sixteen additional clusters in the GCGPS survey resulted in non-detections and currently do not host any known MSPs. For most of these clusters, the imaging reveals several previously unassociated compact radio sources, which could plausibly include yet-undiscovered MSPs. To enable dedicated follow-up observations of these clusters with deeper sensitivity, we publish the processed $\tt{CASA}$ FITS images along with this paper (see DOI: \href{https://doi.org/10.5281/zenodo.17781807}{10.5281/zenodo.17781807}
). We appreciate using these images to identify and prioritise promising targets for future searches.

\subsection{Non-detection limits}
\label{subsec:nondetection}
To estimate the non-detection limit of the undiscovered GC MSPs from the GCGPS observations, we utilise imaging data along with the correlation between continuum emission and the pulsar radiometer equation. This approach allows for a more accurate estimation of detection limits $-$ via the derived $S_\mathrm{min}$ from the GC images $-$ as it inherently accounts for sensitivity losses caused by scintillation or the presence of radio frequency interference (RFI), both of which can lead to sub-optimal phasing.

According to \cite{Dewey_1985}, for continuum emission, the radiometer equation can be written as:
\begin{equation} \label{eqn:continuum_radiometer}
    S_\mathrm{min} = \frac{{SNR} \, * T_\mathrm{sys} \, * \beta}{G \sqrt{N(N-1)n_\mathrm{pol} BW \Delta t_\mathrm{obs}}}
\end{equation}

whereas, for time-domain phased-array (PA) beamforming observation, we can write:
\begin{equation} \label{eqn:pulsar_radiometer}
    S_\mathrm{min} = \frac{{SNR} \, * T_\mathrm{sys} \, * \beta}{G \times N \sqrt{n_\mathrm{pol} BW \Delta t_\mathrm{obs}}}\times\sqrt{\frac{W_\mathrm{eff}}{P_\mathrm{s}-W_\mathrm{eff}}}
\end{equation}

Using Equation \ref{eqn:continuum_radiometer} and Equation \ref{eqn:pulsar_radiometer}, we can write:
\begin{equation} \label{eqn:rms_radiometer}
    S^{p}_\mathrm{min} = S^{c}_\mathrm{min}\times\frac{\sqrt{N_\mathrm{1}(N_\mathrm{1}-1)}}{N_\mathrm{2}}\times\sqrt{\frac{W_\mathrm{eff}}{P_\mathrm{s}-W_\mathrm{eff}}}
\end{equation}

Where all the parameters carry their standard meaning. Here, $\rm{N_\mathrm{1}}$ is the number of antennas used for the continuum observation, and $\rm{N_\mathrm{2}}$ is the number of antennas used for the PA/CD beamforming observation. In GCGPS, for continuum recording, we always used all the available antennas at the time of the observation, which typically ranged from 27 to 30. For beamforming observation, for the central square (CSQ) PA/CD beam, the expected number of antennas is 14 (all the central square antennas), and for the third arm (3rd arm) beam, it's 22 (taking up to the third antenna in each arm for beam formation).

Using Equation~\ref{eqn:rms_radiometer} and the rms value derived from the imaging data ($S^{c}_\mathrm{min}$), we calculate the beam central sensitivity ($S^{p}_\mathrm{min}$) achieved in our beamformed observations. For each GC, we take into account the number of antennas used for the continuum observations ($N_\mathrm{1}$) as well as for the beamforming observations ($N_\mathrm{2}$) while calculating the non-detection limits. For the beamforming non-detection limits, we assume a duty cycle ($\frac{W_{\rm eff}}{P_{\rm s}}$) of 0.1 (10\%), which is a reasonable assumption for the typical MSP population. In Table~\ref{table:observation}, for each GC, we present the $10\sigma$ non-detection limits for both the continuum and the central-square (CSQ) PA/CD beam at the observed band. Similarly, for the third-arm PA/CD beam, we can calculate the achieved beam central sensitivity (non-detection limit) using Equation~\ref{eqn:rms_radiometer}. For the third-arm PA/CD beam, this sensitivity ($S^{p}_\mathrm{3rdarm\_min}$) will scale by a factor of $\sim 1.57$ compared to $S^{p}_\mathrm{CSQ\_min}$, following the relation
$S^{p}_\mathrm{3rdarm\_min} = S^{p}_\mathrm{CSQ\_min} \times \left(\frac{N^\mathrm{CSQ}_\mathrm{2}}{N^\mathrm{3rdarm}_\mathrm{2}}\right)$,
where theoretically $N^\mathrm{CSQ}_\mathrm{2}=14$ and $N^\mathrm{3rdarm}_\mathrm{2}=22$.

\section{SUMMARY}
\label{sec:summary}
This paper presents the discovery of five MSPs from the Globular Clusters GMRT Pulsar Survey (aka GCGPS). All these MSPs are the first MSPs to be discovered in the respective GCs.

The GCGPS observation of the GC NGC~6637 (M69) resulted in the discovery of two MSPs: M69A is a 3.80 ms MSP ($\rm DM\,=\,82.12\,pc\,cm^{-3}$), it was detected with a significant acceleration ($\dot{P}\,\sim\,10^{-12}$) during the discovery observation, confirming it's a binary in a relatively compact orbit; whereas M69B, a 4.80 ms MSP ($\rm DM\,=\,81.92\,pc\,cm^{-3}$), was detected with no acceleration, suggesting it is isolated or a binary system in a wide orbit. We localize both MSPs in the image plane (see Figure \ref{fig:NGC6637_image} and Table \ref{table:MSPdetails}) with arc-sec precision, by simulating the discovery uGMRT beams along with using the simultaneously recorded image data during discovery observation and subsequent follow-up observations.

The observation of NGC~6681 (M70) led to the discovery of three new millisecond pulsars (MSPs). M70A is a 3.93 ms pulsar with a dispersion measure (DM) of 70.65 pc cm$^{-3}$. The brightest among them is M70B, a 4.44 ms MSP with a DM of 71.14 pc cm$^{-3}$, while the faintest is M70C, a 6.06 ms MSP with a DM of 71.06 pc cm$^{-3}$. None of these sources exhibited detectable acceleration over the 2.5-hour discovery observation, indicating they are either isolated or in binary systems with relatively wide orbits. As in the case of M69, the beam simulation, along with the detection ratios of both beams (CSQ and 3rd arm), indicated that all three MSPs are located within the very central region of M70. Imaging (see Figure \ref{fig:NGC6681_image}) revealed a very bright, extended radio source near the center of the globular cluster, which prevented us from confidently associating any of the MSPs with a compact radio source. However, a faint, compact source was also detected almost exactly at the cluster center. This source is a strong candidate for being associated with one of the MSPs. Dedicated timing follow-up is required to determine the precise positions of the MSPs and to confirm any association with the detected radio source.

The imaging of all globular cluster fields observed in the GCGPS survey, including those that did not yield any MSP discoveries (see Figures~\ref{fig:GCGPS_clusters_1}, \ref{fig:GCGPS_clusters_2}, and \ref{fig:GCGPS_clusters_3}), revealed several promising targets that could potentially host undiscovered MSPs. These candidates may have been missed in the original GCGPS search due to reduced sensitivity, particularly in the outer regions of the clusters. The processed FITS files corresponding to these images are published with this paper (DOI: \href{https://doi.org/10.5281/zenodo.17781807}{10.5281/zenodo.17781807}) to support more focused, targeted follow-up efforts. Additionally, the imaging data were used to compute non-detection limits for each of the GCGPS-observed globular clusters using the radiometer equation (Equation~\ref{eqn:rms_radiometer}). Based on the respective observational parameters, the $10\sigma$ non-detection limits (both continuum and beamformed) for each cluster, at the observed uGMRT frequency band, are presented in Table~\ref{table:observation}.

\begin{acknowledgments}
We acknowledge the support of the Department of Atomic Energy, Government of India, under project No. 12-R\&D-TFR5.02-0700. The GMRT is run by the National Centre for Radio Astrophysics of the Tata Institute of Fundamental Research, India. We thank the GMRT operators for their coordinated effort in conducting the GCGPS survey observations. PCCF gratefully acknowledges continuing support from the Max-Planck-Gesellschaft and hospitality of the Academia Sinica Institute of Astronomy and Astrophysics in Taipei, Taiwan, where part of this work was conducted while he was a Visiting Scholar. We also acknowledge support from the \textit{Building Indo-UK Collaborations towards the SKA} program, which facilitated the development of the pulsar search pipeline, particularly the highly efficient GPU-based version. We acknowledge the contributions of the SPOTLIGHT team in the development of the pulsar search pipeline. SPOTLIGHT at the GMRT enables studies of pulsars and fast radio bursts (FRBs) using a petaflop-scale computing facility (called Param Brahmand) based on Param Rudra servers, funded by the National Supercomputing Mission of the Government of India. We especially thank our colleagues at C-DAC (Centre for Development of Advanced Computing) for their support in setting up the Param Brahmand data centre at the GMRT.   
\end{acknowledgments}

\appendix

\clearpage

\section{PRESTO PREPFOLD output of all five discovered MSPs}
\label{appendix:MSP_prepfold_outputs}

\begin{figure*}[h]
\centering
\setlength{\tabcolsep}{2pt}
\renewcommand{\arraystretch}{1.5}
\begin{tabular}{cc}
\includegraphics[width=0.45\textwidth]{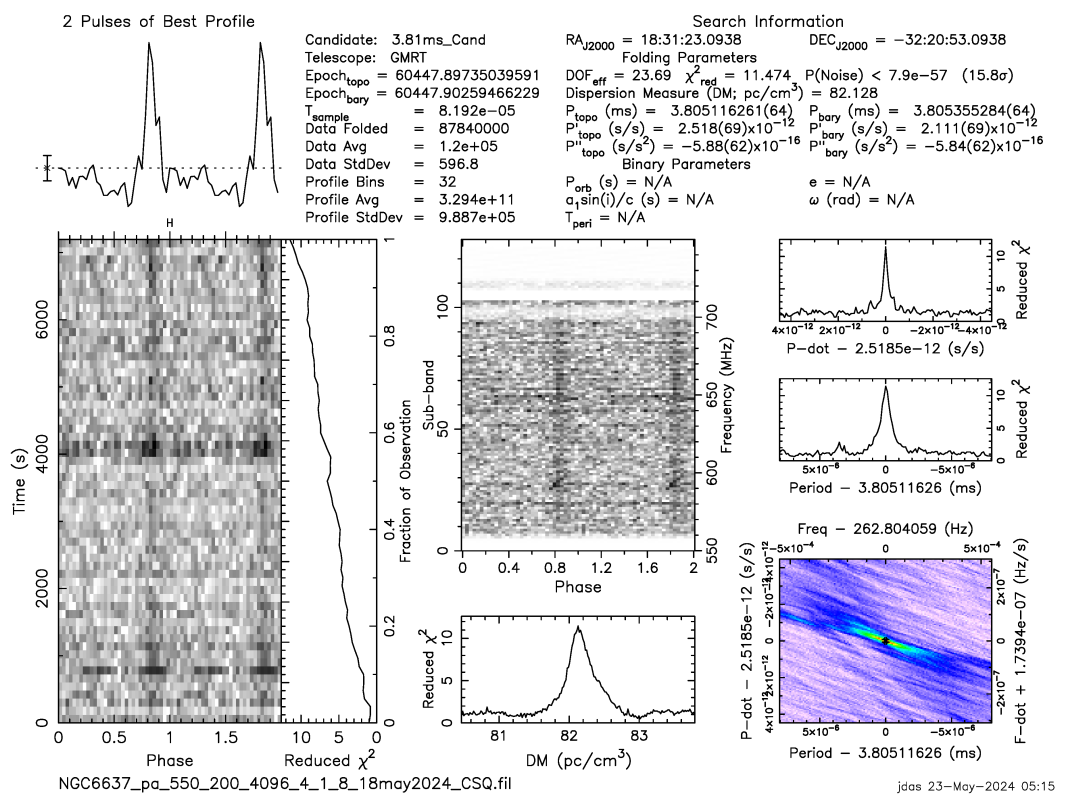} &
\includegraphics[width=0.45\textwidth]{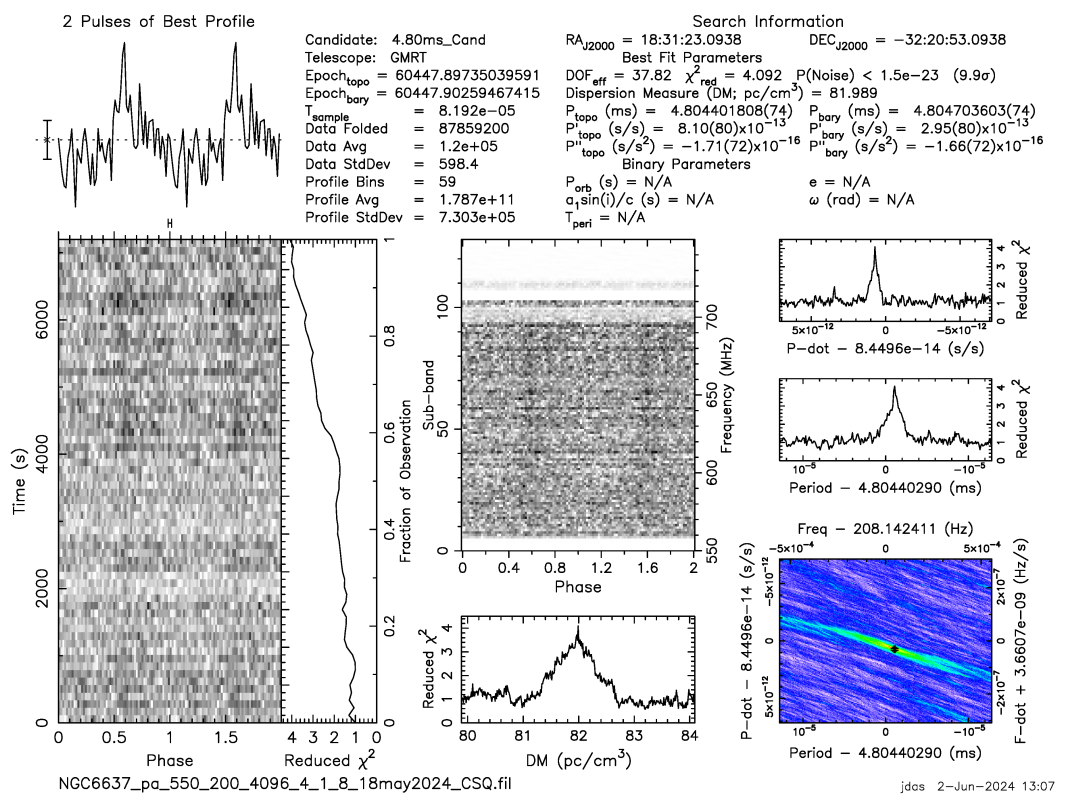} \\
(a) J1831$-$3220A (M69A) & (b) J1831$-$3220B (M69B)  \\
\includegraphics[width=0.45\textwidth]{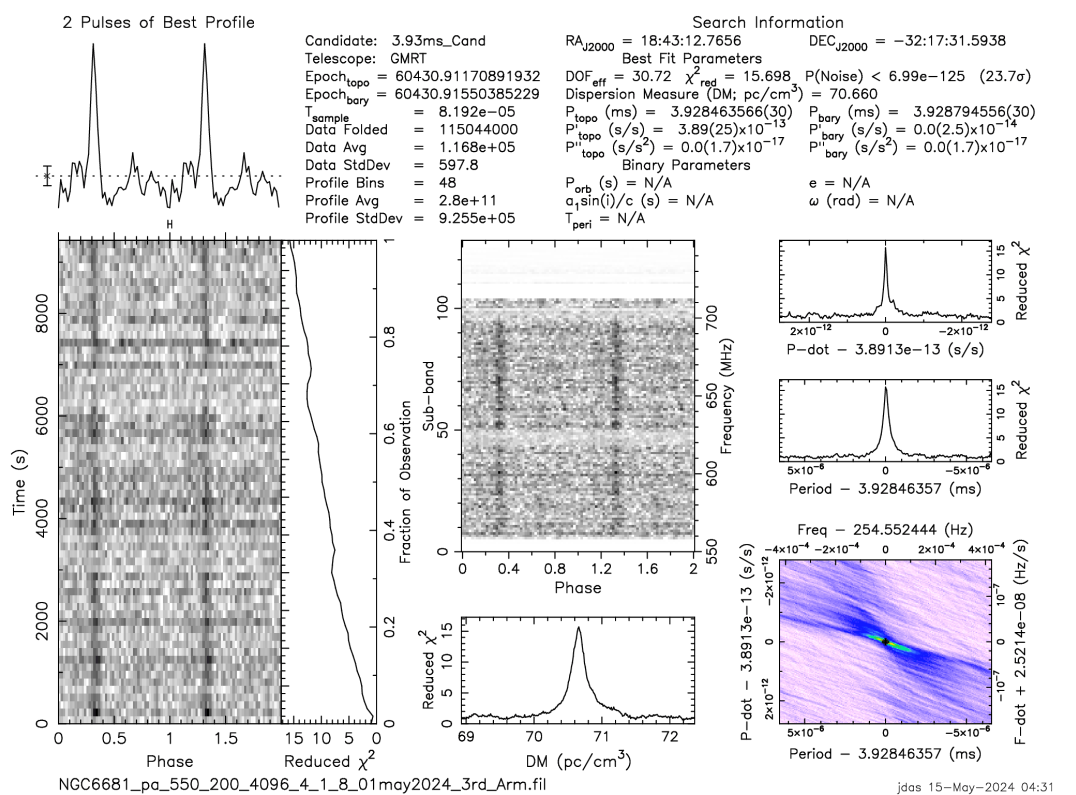} &
\includegraphics[width=0.45\textwidth]{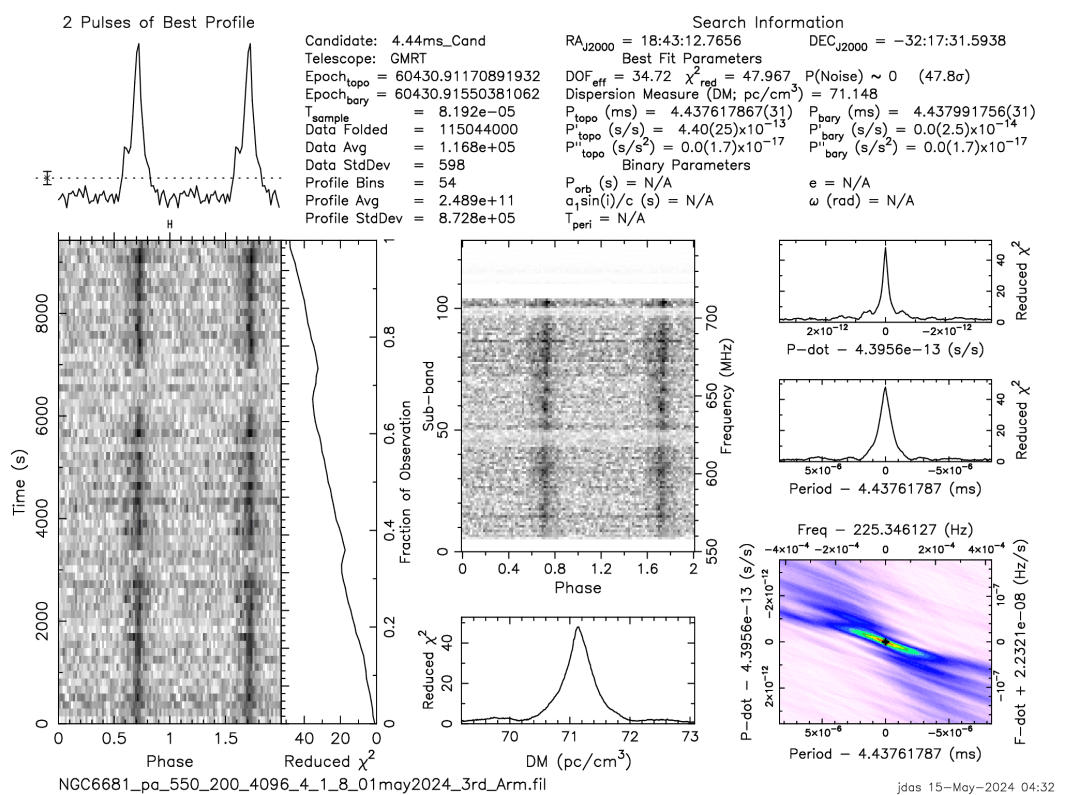} \\
(c) J1843$-$3217A (M70A) & (d) J1843$-$3217B (M70B)  \\
\includegraphics[width=0.45\textwidth]{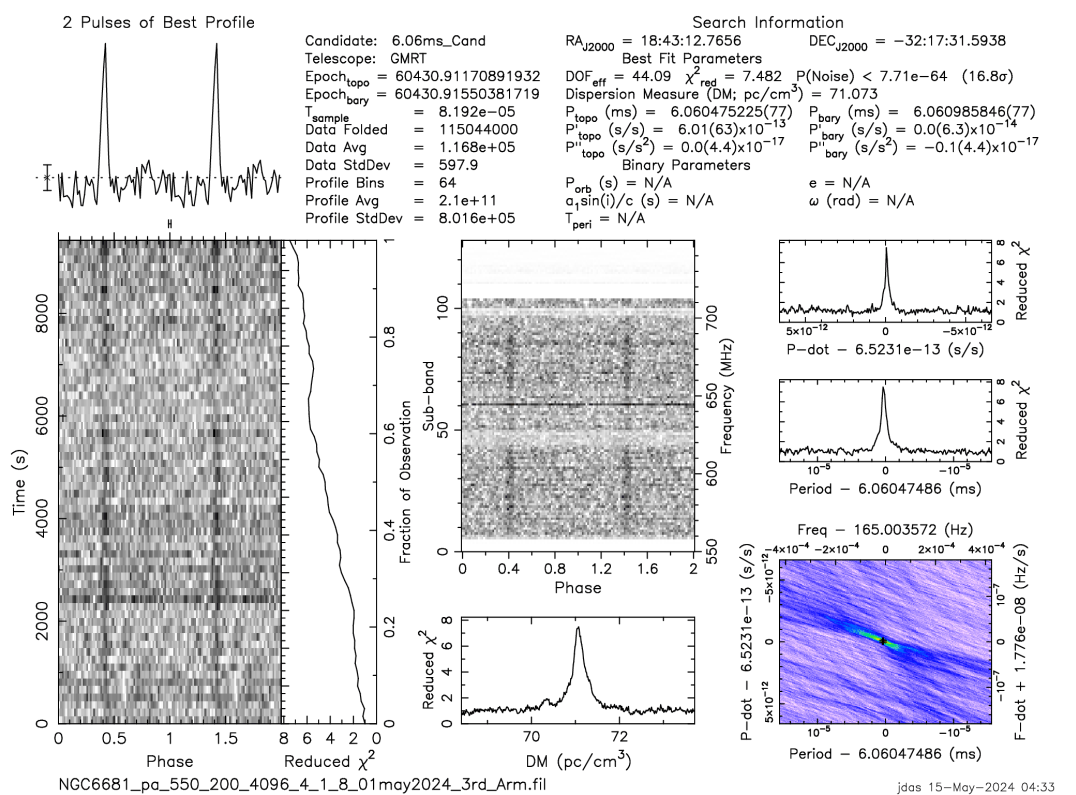} \\
(e) J1843$-$3217C (M70C)  \\
\end{tabular}
\caption{\texttt{PRESTO PREPFOLD} output for the highest-S/N beam from the discovery observations of the five newly discovered MSPs in M69 and M70.}
\label{fig:GCGPS_MSP_folded_profiles}
\end{figure*}

\clearpage
\section{Radio Continuum Images of GCGPS Observed GCs}
\label{appendix:GC_images}

\begin{figure*}[h]
\centering
\setlength{\tabcolsep}{4pt}
\renewcommand{\arraystretch}{2}
\begin{tabular}{ccc}
\includegraphics[width=0.30\textwidth]{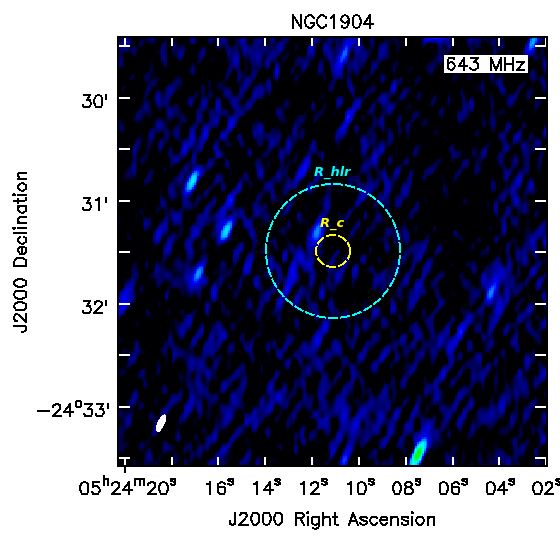} &
\includegraphics[width=0.30\textwidth]{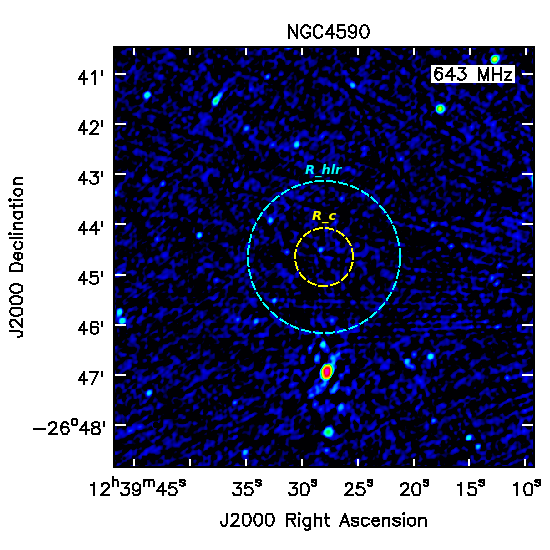} &
\includegraphics[width=0.30\textwidth]{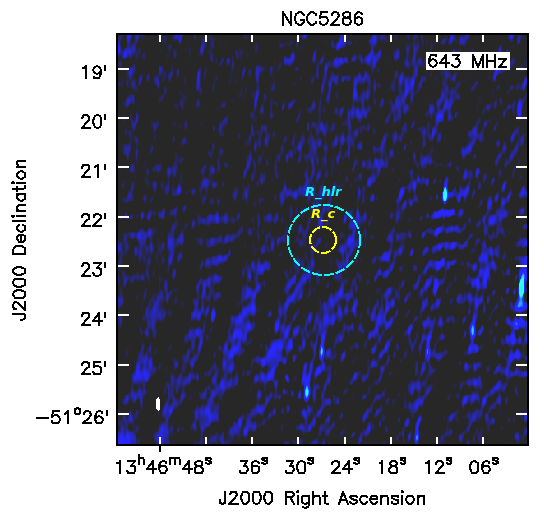} \\
(a) NGC1904 & (b) NGC4590 & (c) NGC5286 \\
\includegraphics[width=0.30\textwidth]{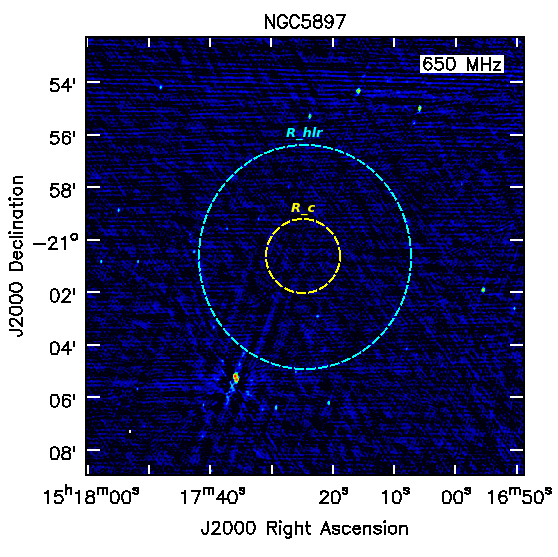} &
\includegraphics[width=0.30\textwidth]{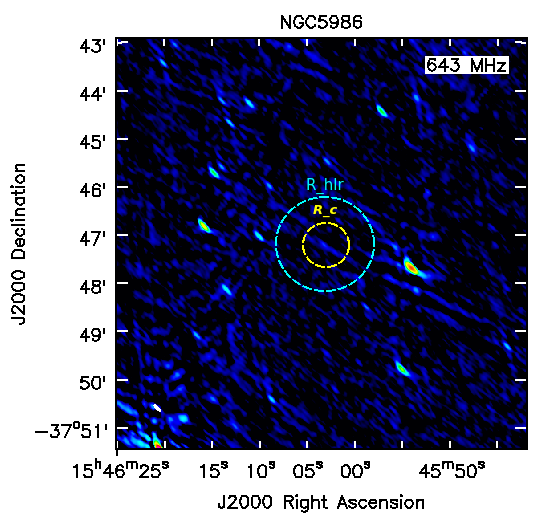} &
\includegraphics[width=0.30\textwidth]{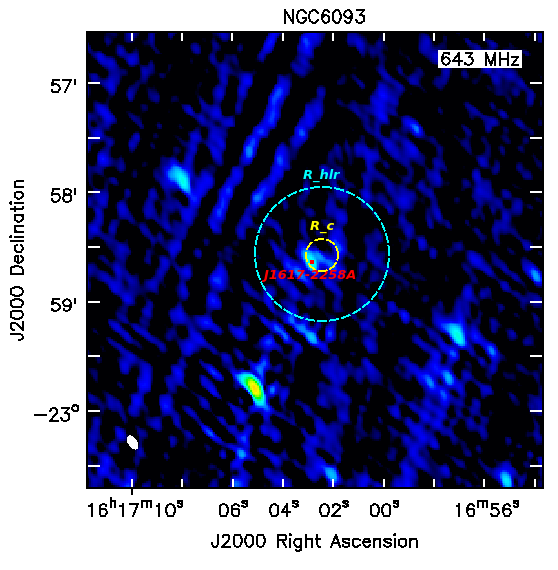} \\
(d) NGC5897 & (e) NGC5986 & (f) NGC6093 \\
\includegraphics[width=0.30\textwidth]{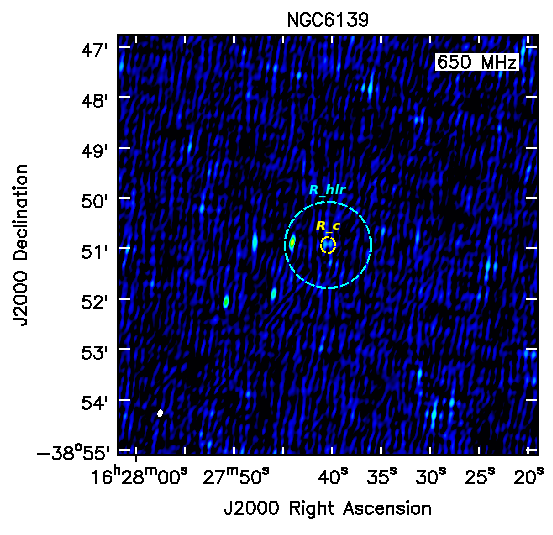} &
\includegraphics[width=0.30\textwidth]{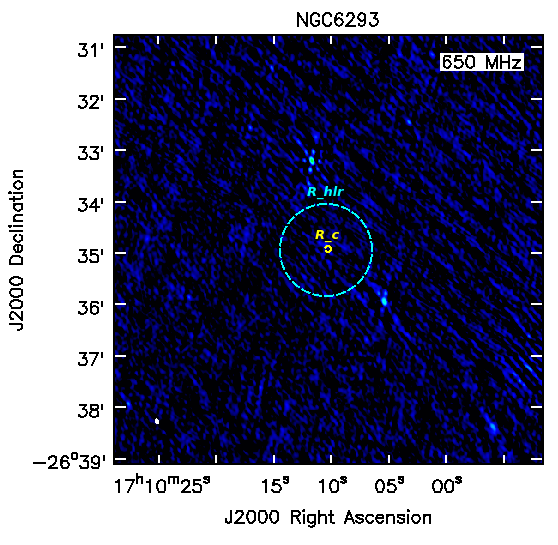} &
\includegraphics[width=0.30\textwidth]{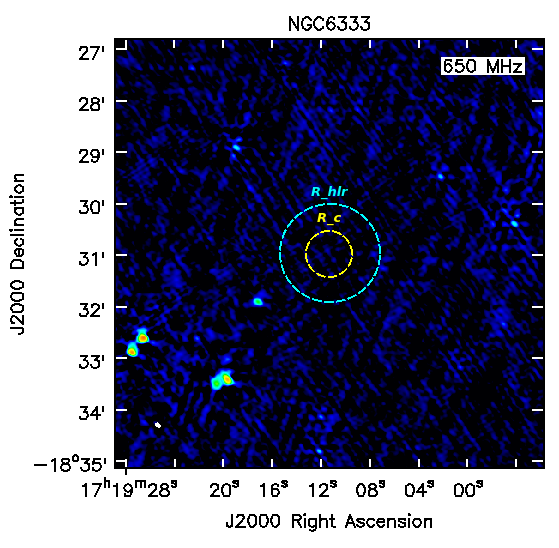} \\
(g) NGC6139 & (h) NGC6293 & (i) NGC6333 \\
\end{tabular}
\caption{Radio continuum images of the globular clusters observed in the GCGPS survey (Set~1). 
In each image (Set~1, as well as Sets~2 and 3 on the following pages), the dashed cyan circle denotes the half-light radius, $\rm{R_{\mathrm{hlr}}}$, and the yellow circle marks the core radius, $\rm{R_{\mathrm{c}}}$, indicating the core region of the cluster.}
\label{fig:GCGPS_clusters_1}
\end{figure*}

\begin{figure*}[p]
\centering
\setlength{\tabcolsep}{4pt}
\renewcommand{\arraystretch}{2}
\begin{tabular}{ccc}
\includegraphics[width=0.30\textwidth]{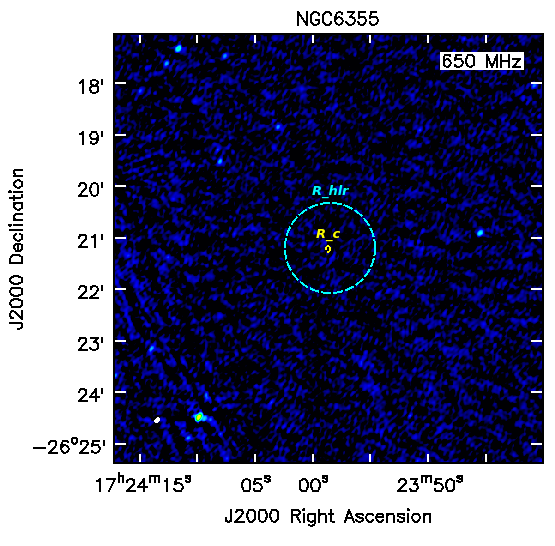} &
\includegraphics[width=0.30\textwidth]{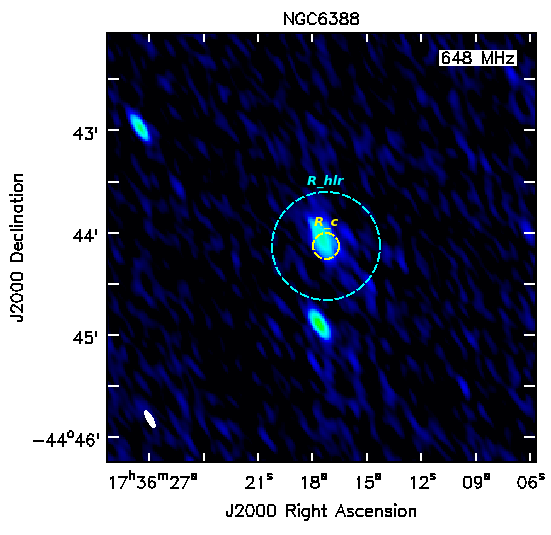} &
\includegraphics[width=0.30\textwidth]{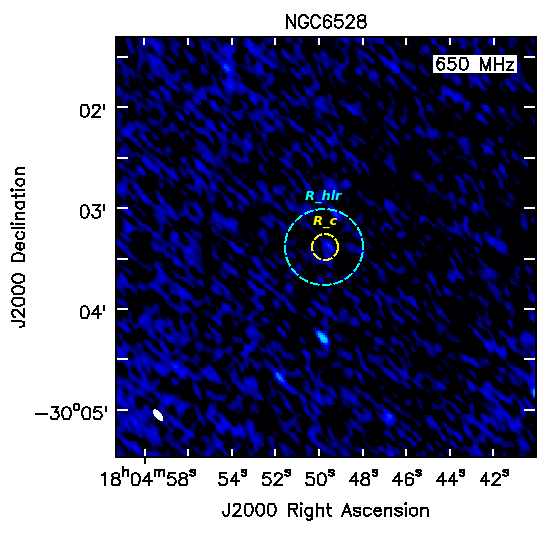} \\
(a) NGC6355 & (b) NGC6388 & (c) NGC6528 \\
\includegraphics[width=0.30\textwidth]{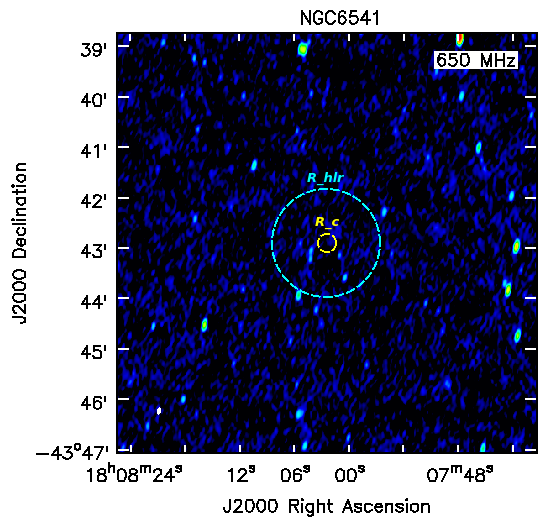} &
\includegraphics[width=0.30\textwidth]{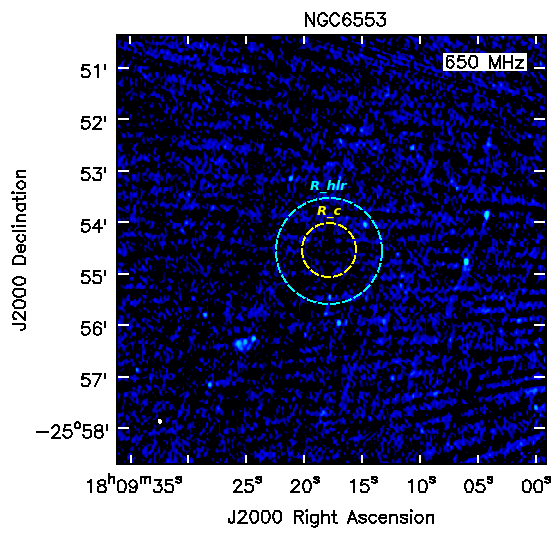} &
\includegraphics[width=0.30\textwidth]{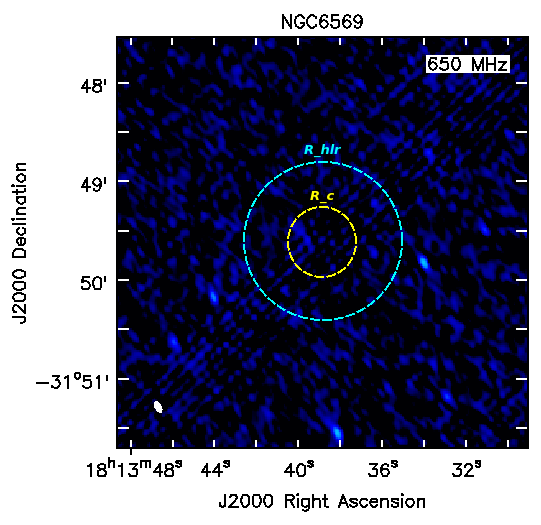} \\
(d) NGC6541 & (e) NGC6553 & (f) NGC6569 \\
\includegraphics[width=0.30\textwidth]{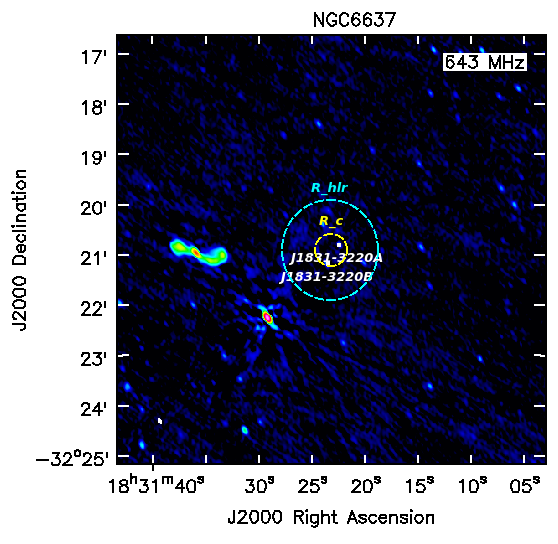} &
\includegraphics[width=0.30\textwidth]{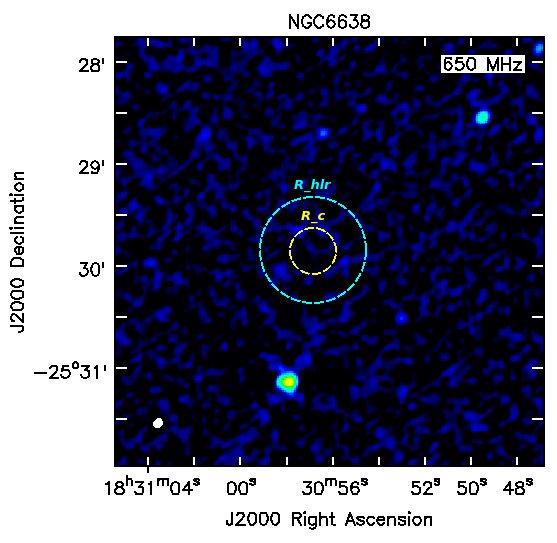} &
\includegraphics[width=0.30\textwidth]{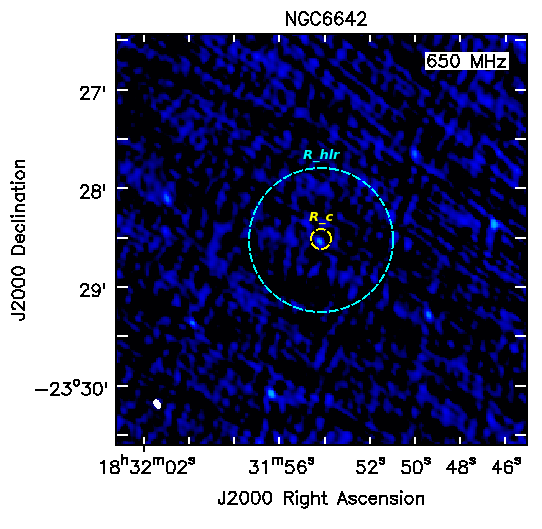} \\
(g) NGC6637 & (h) NGC6638 & (i) NGC6642 \\
\end{tabular}
\caption{Radio continuum images of GCs observed in the GCGPS survey (Set 2).}
\label{fig:GCGPS_clusters_2}
\end{figure*}

\begin{figure*}[p]
\centering
\setlength{\tabcolsep}{4pt}
\renewcommand{\arraystretch}{2}
\begin{tabular}{ccc}
\includegraphics[width=0.30\textwidth]{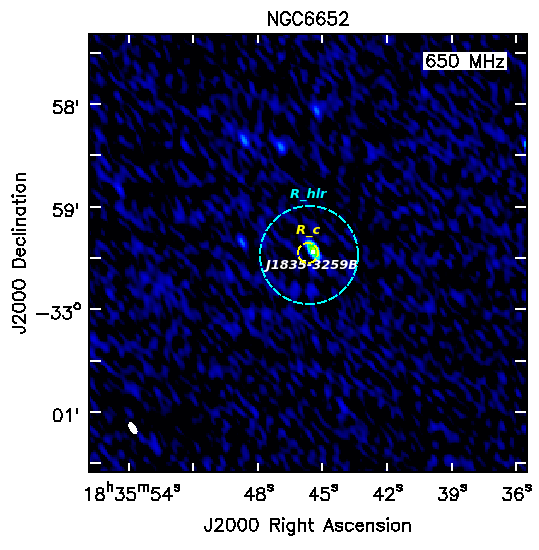} &
\includegraphics[width=0.30\textwidth]{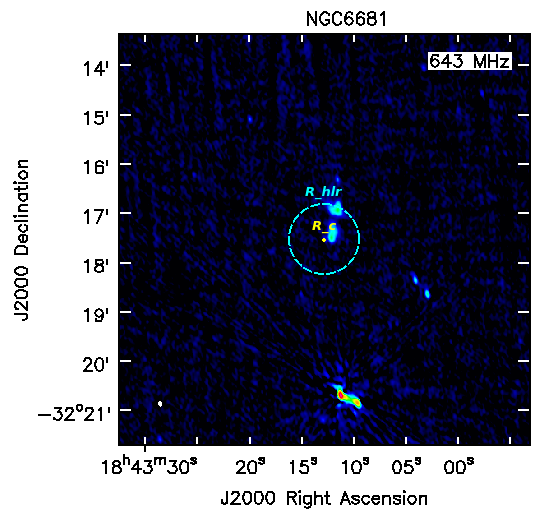} &
\includegraphics[width=0.30\textwidth]{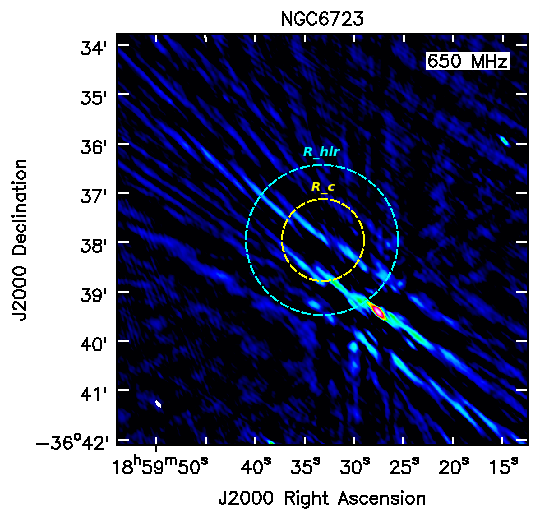}\\
(a) NGC6652 & (b) NGC6681 & (c) NGC6723 \\
\includegraphics[width=0.30\textwidth]{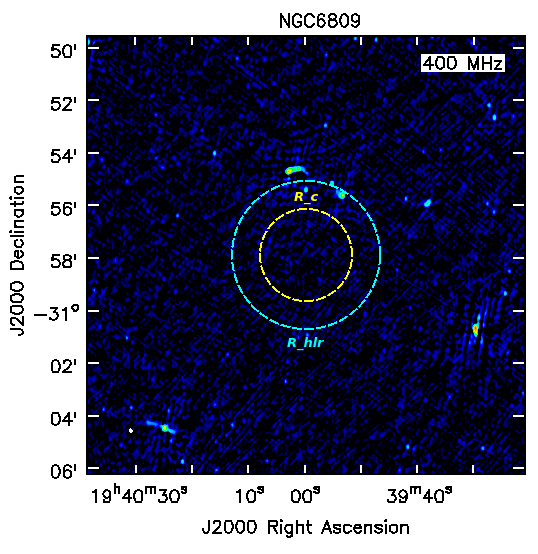} \\
(d) NGC6809 \\
\end{tabular}
\caption{Radio continuum images of GCs observed in the GCGPS survey (Set 3).}
\label{fig:GCGPS_clusters_3}
\end{figure*}

\clearpage

\bibliography{bibliography}{}
\bibliographystyle{aasjournal}

\end{document}